\documentclass[prl,reprint,amsmath,nofootinbib,superscriptaddress,showpacs]{revtex4-1}
\usepackage{amssymb}
\usepackage{empheq}

\usepackage[T1]{fontenc} 

\usepackage[utf8]{inputenc}  
\usepackage{slashed,epsfig,amssymb,bm,graphicx,amsmath,amsfonts}
\usepackage{mathtools}
\usepackage{setspace}
\usepackage{mathrsfs} 
\usepackage[colorlinks]{hyperref}
\usepackage{graphicx}
\usepackage{empheq} 
\usepackage{color}
\usepackage{ulem}
\newcommand{\be}{\begin{equation}}
\newcommand{\ee}{\end{equation}}
\newcommand{\beq}{\begin{equation}}
\newcommand{\eeq}{\end{equation}}
\newcommand{\bea}{\begin{eqnarray}}
\newcommand{\eea}{\end{eqnarray}}

\newcommand{\nn}{\nonumber}
\newcommand{\llll}{\langle\langle}
\newcommand{\rr}{\rangle\rangle}

\def\k{{\bf k}}

\def\[{\left[}
\def\]{\right]}
\def\({\left(}
\def\){\right)}
\def\<{\langle}
\def\>{\rangle}
\def\O{\mathcal{O}}
\def\C{\mathcal{C}}
\def\L{\mathcal{L}}

\def\Mp{M_{\rm p}}

\def\rarr{\rightarrow}

\begin{document}

\begin{abstract}

We argue that massive quantum fields source low-frequency long-wavelength metric fluctuations through the quantum fluctuations of their stress-energy, given reasonable assumptions about the analytic structure of its correlators. This can be traced back to the non-local nature of the gauge symmetry in General Relativity, which prevents an efficient screening of UV scales (what we call the cosmological non-constant problem). We define a covariant and gauge-invariant observable which probes line-of-sight spacetime curvature fluctuations on an observer's past lightcone, and show that current pulsar timing data constrains any massive particle to $m\lesssim 600$ GeV. This astrophysical bound severely limits the possibilities for physics beyond the standard model below the scale of quantum gravity. 
\end{abstract} 

\title{Pulsar Timing Constraints on Physics Beyond the Standard Model}

\author{Niayesh Afshordi}
\email{nafshordi@pitp.ca}
\affiliation{Perimeter Institute for Theoretical Physics, 31 Caroline St. N., Waterloo, ON, N2L 2Y5, Canada}
\affiliation{Department of Physics and Astronomy, University of Waterloo, Waterloo, ON, N2L 3G1, Canada}

\author{Hyungjin Kim}
\email{h268kim@uwaterloo.ca}
\affiliation{Department of Applied Mathematics, University of Waterloo, Waterloo,
Ontario, N2L 3G1, Canada}

\author{Elliot Nelson}
\email{enelson@pitp.ca}
\affiliation{Perimeter Institute for Theoretical Physics, 31 Caroline St. N., Waterloo, ON, N2L 2Y5, Canada}

\date{\today}    

\maketitle

General Relativity (GR) couples geometry to classical stress-energy via Einstein's field equations. Semiclassically, the leading gravitational effect of quantum fields is through the expectation value $\<T_{\mu\nu}\>$, the apparent divergence of which leads to the cosmological constant (CC) problem \cite{Straumann:2002tv, Weinberg:1988cp}. Furthermore, quantum operators generically have some nonzero spread around their expected values, which can lead to additional gravitational effects from two-point and higher functions:
\be\label{Einstein_2pt}
\<G_{\mu\nu}G_{\alpha\beta}\>_{c,\rm vac} = \Mp^{-4} \<T_{\mu\nu}T_{\alpha\beta}\>_{c,\rm vac},
\ee
where we have explicitly denoted the connected contribution from zero-point vacuum fluctuations, and the LHS is the contribution to Einstein tensor fluctuations from matter in the vacuum. Such gravitational effects from vacuum fluctuations have been reviewed in \cite{Ford:2007gb} (see also \cite{Carlip:2015ana,Wang:2017oiy}), and in the stochastic gravity formalism \cite{Hu:2008rga}.

In \cite{Afshordi:2015iza}, we pointed out that there is no natural mechanism in GR to decouple UV fluctuations of vacuum stress-energy from IR geometry. As such, cosmological observables are sensitive to the UV scales in the vacuum of quantum fields, which we dubbed the {\it cosmological non-constant  (CnC) problem}. What makes GR different from other quantum field theories is the non-local nature of its gauge symmetry, leading to a failure of natural screening mechanisms of UV physics: Particles and anti-particles have the same gravitational charge and thus cannot screen each other, while conservation and positivity of energy is not guaranteed\footnote{Unless the spacetime is asymptotically flat, which as we shall see is precluded in solutions of (\ref{Einstein_2pt}).}, allowing contributions of UV modes to lR processes. We then argued that the classical notion of geometry is only valid locally, while its quantum fluctuations $h$ diverge in the IR as:
\be\label{CnC}
h^2 \sim \frac{m^5 \times ({\rm Length ~or~Time})}{\Mp^4},
\ee
where $m$ is the highest UV scale in the theory.

In this {\it letter}, we study the gravitational effect of vacuum stress-energy fluctuations of massive quantum fields using pulsar timing data, which probes the left hand side of Eq. \eqref{Einstein_2pt} with high precision:
We first identify a \textit{negative} low-frequency part of the stress-energy two-point function of massive fields, using an analytic deformation in the complex frequency plane.
We then compute a covariant and gauge-invariant correlation function of the Riemann curvature which determines the change in frequency of a photon moving through a perturbed geometry inbetween two timelike geodesics (e.g., from pulsar to Earth).
The power spectrum of fluctuations in the period of observed pulses can then be related to the Riemann curvature fluctuations which are sourced by the vacuum stress-energy fluctuations.
We obtain a result, Eq. \eqref{pulsar_CnC}, of the same form as Eq. \eqref{CnC}.
Finally, we compare this result with pulsar timing data to put an upper bound (subject to some disclaimers) on the mass of quantum fields, which severely constrains particle physics beyond the standard model (BSM).


We use the $-+++$ signature convention, denote four-momentum as $k$, and three-momentum as $\k$.

\textit{Stress-energy correlators as $k^2\rarr0$.}
A Lorentz-invariant stress-energy tensor two-point function takes the general form\footnote{Note that we do not time-order. The $_s$ subscript denotes symmetrization, e.g., $\langle XY\rangle_s=(\langle XY\rangle+\langle YX\rangle)/2$. The $_c$ footnote denotes the connected part, obtained after subtracting off the disconnected part $\langle T_{\mu\nu}\rangle\langle T_{\alpha\beta}\rangle$.}
\bea
&&\llll T_{\mu\nu}(k)T_{\alpha\beta}(k')\rr_{s,c} = 2\pi\Big[\rho_0(-k^2)P_{\mu\nu}P_{\alpha\beta} + \nn \\
&& \ \ \ \rho_2(-k^2)\Big(\frac{1}{2}P_{\mu\alpha}P_{\nu\beta}+\frac{1}{2}P_{\mu\beta}P_{\nu\alpha}-\frac{1}{3}P_{\mu\nu}P_{\alpha\beta}\Big)\Big], \label{TTgeneral}
\eea
where the projection tensors, $P_{\mu\nu}=\eta_{\mu\nu}-k_\mu k_\nu/k^2$, ensure that $\partial^\mu T_{\mu\nu}=0$, and the double expectation value indicates removal of an overall factor of $(2\pi)^4\delta^4(k+k')$. The spin-0 and spin-2 spectral densities $\rho_{0,2}$ are non-negative \cite{Afshordi:2015iza}.

In \cite{Afshordi:2015iza} (see also \cite{Hu:2008rga,Martin:2000dda}) we computed $\rho_{0,2}$ for a massive scalar field with Lagrangian $\mathcal{L}_\phi = -\frac{1}{2}[(\partial\phi)^2+m^2\phi^2]$:
\bea
&& \hspace{-0.3cm} \rho^{(\phi)}_0(-k^2) = \frac{k^4}{144\pi^2}\sqrt{1+4\frac{m^2}{k^2}}\left[\frac{1}{2}-\frac{m^2}{k^2}\right]^2 \Theta(-k^2-4m^2), \nn \\
&& \rho^{(\phi)}_2(-k^2) = \frac{k^4}{1920\pi^2}\Big(1+\frac{4m^2}{k^2}\Big)^{5/2} \Theta(-k^2-4m^2), \label{rho_phi}
\eea
Note that in the $k^2\rightarrow0$ limit (ignoring the $\Theta$ function), we have $\rho^{(\phi)}_2=(6/5)\rho^{(\phi)}_0$. As discussed in \cite{Afshordi:2015iza}, this implies the complete symmetry of $\langle T_{\mu\nu}T_{\alpha\beta}\rangle$ in its indices, for small $k^2$, which can be understood in terms of a Poisson distribution in the phase space.

For a Dirac field, $\mathcal{L}_{\rm Dirac} = \bar{\psi}(i\gamma^\mu\partial_\mu - m)\psi$, the trace of the stress tensor appearing in Einstein's equations, $T_{\mu\nu}=\frac{2}{\sqrt{-g}}\frac{\delta S}{\delta g^{\mu\nu}}$, has the simple form $(T_\mu^\mu)^{(\psi)}=m\bar{\psi}\psi$. Contracting Eq. \eqref{TTgeneral}, we see that $\langle T^\mu_\mu T^\nu_\nu\rangle$ is proportional to $\rho_0$. Consequently, evaluating the two-point function we find
\bea
\rho^{(\psi)}_0(-k^2) &=& \frac{1}{144\pi^2} m^2 (k^2+4m^2) \Big(1+\frac{4m^2}{k^2}\Big)^{1/2} \nn \\
&& \hspace{2cm} \times \Theta(-k^2-4m^2). \label{rho0_psi}
\eea
We will be interested in the $k^2\rightarrow 0$ regime, in which case we will again have $\rho_2^{(\psi)}=(6/5)\rho_0^{(\psi)}$.

For a real massive spin-1 field, using $\L_A=-\frac{1}{4}F_{\mu\nu}^2-\frac{1}{2}m^2 A_\mu^2$, we obtain $T_\mu^\mu=-m^2A_\mu A^\mu$, which leads to
\bea
\rho^{(A)}_0(-k^2) &=& \frac{1}{144\pi^2} (3m^4+m^2k^2+\frac{1}{4}k^4) \Big(1+\frac{4m^2}{k^2}\Big)^{1/2} \nn \\
&& \hspace{2cm} \times \Theta(-k^2-4m^2). \label{rho0_A}
\eea

\begin{figure}
\begin{center}
\includegraphics[scale=0.45]{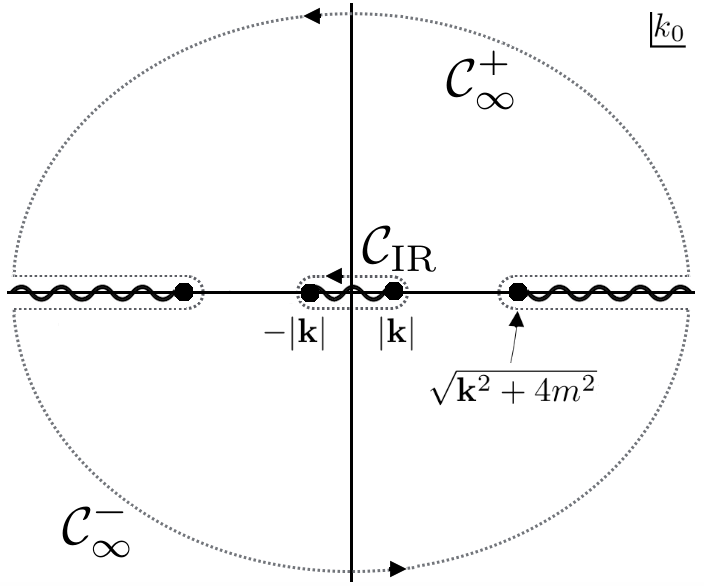}
\end{center}
\caption{The frequency integral (when transforming stress-energy correlators to real space) can be deformed in the complex plane as expressed in Eq. \eqref{contour}, with the contours at infinity $\C_\infty^\pm$ and low-frequency contour $\C_{\rm IR}$ as shown. The branch cuts of the spectral densities may be chosen as indicated here.}
\label{fig:contour}
\end{figure}

Stress-energy tensor correlations in real space are obtained by Fourier transforming Eq. \eqref{TTgeneral}. The frequency integral, which is restricted to $\omega^2>4m^2+|\k|^2$, can be deformed in the complex plane:
\beq\label{contour}
\hspace{-0.35cm}
\int_{-\infty}^\infty dk_0 \Theta(-k^2-4m^2)\times... = \frac{1}{2}\(\int_{\C_\infty^+}  + \int_{\C_\infty^-} + \int_{\C_{\rm IR}} \) dk_0 \times...,
\eeq
where $\C_\infty^\pm$ and $\C_{\rm IR}$ are shown in Figure \ref{fig:contour}.


Restricting the integration on the real line to $k_0^2<\Lambda^2$, and letting $\Lambda\rarr\infty$, the contours at infinity $C_\infty^\pm$ involve power law divergences in $\Lambda$ (with only odd powers), which we expect to be absent after renormalization.  The low-frequency contour $\C_{\rm IR}$, on the other hand, is UV convergent, so we may effectively replace the step function in Eqs. \eqref{rho_phi}-\eqref{rho0_A} with $-\Theta(k^2)$, which picks out $\frac{1}{2}\int_{\C_{\rm IR}}$. This allows us to use the low frequency spectral density
\be
\rho^{(\phi)}_{0,\rm IR}(-k^2) = - \frac{1}{72\pi^2}\frac{m^5}{\sqrt{k^2}}\Theta(k^2),
\ee
along with $\rho^{(\psi)}_{0,\rm IR}=4\rho^{(\phi)}_{0,\rm IR}$ and  $\rho^{(A)}_{0,\rm IR}=3\rho^{(\phi)}_{0,\rm IR}$, and $\rho^{(X)}_{2,\rm IR}=\frac{6}{5}\rho^{(X)}_{0,\rm IR}$, for each species.
The negative sign here is obviously significant, and will be discussed below.

Finally, we note that a complex scalar or vector field will have an extra factor of two relative to the real case, and that a Majorana field will have a factor of half relative to the Dirac case.

\textit{Covariant Pulsar Timing.}
We start by deriving a covariant formula for pulsar timing observations, to 1st order in metric perturbations. Imagine a continuous family of time-like non-intersecting geodesics which includes the Earth and pulsar worldlines. Assuming that each geodesic characterizes a fixed spatial coordinate, $x^i$, and taking the proper time along the geodesic as time coordinate, $t$, the metric will have the form:
\beq
ds^2= -dt^2+\gamma_{ij} dx^i dx^j,
\eeq
which is also known as the {\it synchronous comoving} gauge.

The energy of a photon that moves from $x^i$ to $x^i+ \Delta x^i$ in these coordinates changes by a parallel transport:
\beq
\Delta E = -\Gamma^0_{\mu\nu} p^\mu dx^\nu = -\frac{1}{2}  \dot{\gamma}_{ij} p^i dx^j.
\eeq   
As $\Delta E$ involves first derivatives of metric, it cannot be written as a local covariant form. However, taking its  time derivative and making use of Eq. \eqref{Riemann} below yields a local covariant form:
\beq
\frac{\partial \Delta E}{\partial t} = -\frac{1}{2}  \ddot{\gamma}_{ij} p^i dx^j = R_{i0j0} p^i dx^j = R_{\mu\nu\alpha\beta} u^\nu u^\beta p^\mu dx^\alpha,
\eeq
This is the unique covariant generalization which reduces to the LHS when the 4-velocity of the pulsar relative to the Earth is $u^\mu=\delta_0^\mu$.\footnote{This can be shown using the (a)symmetry properties of the Riemann tensor.}
Now, summing over $\frac{\partial \Delta E}{\partial t}$ for all the neighboring geodesics that interpolate between Earth and the pulsar worldlines, we find a manifestly covariant expression for the time derivative of the photon's energy/frequency in Earth's rest frame (assuming that the pulsar is a standard clock):
\beq
\frac{d E_{\rm obs.}}{d t} = \int_{\rm pulsar}^{\rm Earth}  R_{\mu\nu\alpha\beta} u^\nu u^\beta p^\mu dx^\alpha,
\eeq
where the integral is along the photon trajectory from the pulsar to Earth. Note that this derivation already includes the boundary terms (i.e. Doppler and Sachs-Wolfe terms), by construction. 

Since the Riemann tensor is already first order in curvature, we can use the Minkowski metric to compute the photon trajectory and treat the momentum $p^\mu = E dx^\mu/dt$ as constant. Likewise, we assume $u^\mu = \delta^\mu_0$ at this order, i.e. ignore the motion of Earth and pulsar. Therefore, in terms of Minkowski coordinates, we have:
\bea
&& \frac{d \ln \nu(t) }{d t} = - \frac{d \ln P(t) }{d t} = u^\nu u^\sigma \frac{d x^\mu}{dt} \frac{dx^\rho}{dt}   \label{lnP} \\
&& \times \int^t_{t-L} \hspace{-0.2cm} dt' ~R_{\mu\nu\rho\sigma}\left[\left(t- t' \over L\right) x^i_{\rm pulsar} + \left(t'-t+L  \over L\right) x^i_{\rm Earth}, t' \right], \nn 
\eea
where $\nu$ and $P$ are the observed frequency and period of the pulsar, respectively, while $L$ is the distance between pulsar and Earth.

The pulsar frequency fluctuations depend on metric fluctuations, $h_{\mu\nu} = g_{\mu\nu} - \eta_{\mu\nu}$, via the the linearized Riemann tensor,\footnote{The Riemann tensor is invariant under linear diffeomphisms around Minkowski spacetime, $h_{\mu\nu} \rightarrow h_{\mu\nu} + \xi_{\mu,\nu}+\xi_{\nu,\mu}$ for arbitrary $\xi_\mu$, and thus is a gauge-invariant observable.}
\beq\label{Riemann}
R_{\mu\nu\rho\sigma} = \frac{1}{2} \left(\partial_\rho\partial_\nu h_{\mu\sigma}+\partial_\sigma\partial_\mu h_{\nu\rho} - \partial_\sigma\partial_\nu h_{\mu\rho} - \partial_\rho\partial_\mu h_{\nu\sigma} \right).
\eeq
We can now use the linearized Einstein equations in the Lorentz gauge\footnote{It is straightforward to show that the dependence of $R_{\mu\nu\rho\sigma}$ on the stress tensor does not depend on the choice of gauge, and can be reproduced by working in conformal Newtonian gauge, for example.} $\Box \bar{h}_{\mu\nu} = 2M^{-2}_p T_{\mu\nu}$, with $\bar{h}_{\mu\nu} = h_{\mu\nu} - \frac{1}{2} h \eta_{\mu\nu}$ to relate the Riemann tensor to the components of the stress tensor.
It will be useful to define the contracted curvature
\beq\label{Rbar}
\bar{R}\equiv u^\nu u^\sigma n^\mu n^\rho R_{\mu\nu\rho\sigma}
\eeq
which appears in Eq. \eqref{lnP}; here, $n^\mu=dx^\mu/dt$. Its two-point function can be evaluated in momentum space using Eq. \eqref{TTgeneral}, with the result:
\bea
\llll\bar{R}_k\bar{R}_{k'}\rr &=& \frac{\pi}{2M_p^4}\frac{1}{k^4}(k\cdot n)^2 \left[ k\cdot n + 2(k\cdot u)(n\cdot u)\right]^2 \nn \\
&& \times \Big(\rho_0(-k^2)+\frac{8}{3}\rho_2(-k^2) \Big). \label{P_R}
\eea

\textit{Power Spectrum.}
Let us next compute the power spectrum of pulsar period, defined as:
\beq
\hspace{-0.25cm}
{\cal P}(\omega) \equiv \omega^{-2}\int d(\Delta t) \left\langle  \frac{d \ln P(t) }{d t} \frac{d \ln P(t') }{d t'} \right\rangle e^{i\omega\Delta t},
\eeq
where $\Delta t=t-t'$. To compare to the pulsar timing literature, note that ${\cal P}(\omega)$ is related to the power spectrum of the timing noise $\Phi_{TN}(f)$ and the (equivalent) amplitude of stochastic gravitational waves, $h_{c,{\rm eq}}$ via:
\beq
{\cal P}(\omega) = (Pf)^{2} \Phi_{TN}(f) = \frac{h^2_{c,{\rm eq}}}{12\pi^2 f},
\eeq
where $f= \omega/(2\pi)$ is the linear frequency (e.g., \cite{Lasky:2015zpa}).

Putting the Earth at the origin and pulsar on the z-axis, that is $n_\mu=(-1,-\hat{z})$, and substituting Eq. (\ref{lnP}) and Eq. \eqref{Rbar} into this definition gives
\beq \label{hc_om}
h^2_{c,{\rm eq}}= \frac{12\pi}{\omega} \int \frac{d^3\k}{(2\pi)^3} \left\{\frac{1-\cos\left[(\omega+k_z)L\right]}{(\omega+k_z)^2}\right\}
\llll R_k R_{k'}\rr,
\eeq 
where $k^2=-\omega^2+k_z^2+k_\perp^2$.
Setting $n_\mu=(-1,-\hat{z})$ and $u_\mu=(-1,\mathbf{0})$, Eq. \eqref{P_R} takes the simple form
\beq\label{RRfinal}
\hspace{-0.2cm}
\llll R_k R_{k'}\rr = \frac{\pi}{2M_p^4} \frac{(k_z^2-\omega^2)^2}{k^4} \Big( \rho_0(-k^2)+\frac{8}{3}\rho_2(-k^2) \Big).
\eeq
Following the contour deformation to low frequencies (Figure \ref{fig:contour}), we use the effective low frequency spectral density,
\bea
\rho^{(X)}_{\rm eff}(-k^2) &=& \rho^{(X)}_{0,{\rm{IR}}}(-k^2) + \frac{8}{3} \rho^{(X)}_{2,{\rm{IR}}}(-k^2) \nn \\
&=& - c_X\frac{7}{120\pi^2}\frac{m^5}{\sqrt{k^2}} \Theta(k^2), \label{rho_eff}
\eea
where
\bea
c_X &=& 1 \ \ \ (X=\phi, \text{ real}) \nn \\
&=& 2 \ \ \ (X=\phi, \text{ complex}) \nn \\
&=& 2 \ \ \ (X=\psi, \text{ Majorana}) \nn \\
&=& 4 \ \ \ (X=\psi, \text{ Dirac}) \nn \\
&=& 3 \ \ \ (X=A_\mu, \text{ real}) \nn \\
&=& 6 \ \ \ (X=A_\mu, \text{ complex}).
\eea

Evaluating the integral over $k_\perp$ in Eq. \eqref{hc_om}, and using Eqs. \eqref{RRfinal}-\eqref{rho_eff}, we find
\bea
\hspace{-0.75cm}
(h^2_{c,{\rm eq}})^{(X)} &=& - c_X\frac{7}{480\pi^3} \frac{m^5}{M_p^4\omega}\int dk_z \frac{|k_z-\omega |^{1/2}}{|k_z+\omega |^{3/2}} \nn \\
&& \times [1-\cos((k_z+\omega)L)] \Theta(k_z^2-\omega^2),
\eea
with the step function resulting from a Gamma function regularization which sets the (divergent) $k_z^2<\omega^2$ contribution to zero.
Finally, integrating over $k_z$ with a cutoff $|k_z|<k_{\rm max}$, we find:
\bea
&& (h^2_{c,{\rm eq}})^{(X)}
\approx - c_X\frac{7}{480\pi^3} \frac{m^5}{M_p^4\omega} \left[ \sqrt{4\pi\omega L} + \ln (k_{\rm max} L) \right] \nn \\
&& \hspace{0.4cm} \approx - 4\times 10^{-30} c_X \(\frac{m}{600 \ \rm GeV}\)^5
\sqrt{\frac{2\pi L({\rm kpc})}{\omega({\rm yr}^{-1})}} \nn \\
\label{pulsar_CnC}
\eea
where we have dropped $\O(1)$ contributions within the brackets. Since we are using the geodesic equation to compute the pulse propagation, a typical pulse size would define $k_{\rm max} c \sim {\rm \mu s}^{-1}$ for the relevant metric perturbations. Therefore, for $L \sim $ kpc and $f \sim$ yr$^{-1}$ the log-corrections are less than 20\%.      

\textit{Constraints on BSM.}
Current pulsar timing data constrains any variation in the period between pulses to be extremely small. The strongest current bound on $h_{c,{\rm eq}}$ comes from the timing noise of PSR J1909-3744 (at $L= 1.23 \pm 0.05$ kpc) \cite{Shannon:2015ect}, at the lowest observed frequency:
\be
h^{\rm (total)}_{c,{\rm eq}}(2\pi\times 0.195 ~{\rm yr}^{-1}) < 3.20\times 10^{-15} ~~(95\%~{\rm C.L.}).
\ee
This constrains the sum of the CnC contribution and other contributions from internal UV physics of the pulsar. If the negative CnC contribution contributes to a sensible observable variance of fluctuations, the sum total must be positive. Making the mild assumption that there is not a fine-tuned cancellation between an extremely large unknown contribution from UV physics, and a large negative contribution from vacuum fluctuations, we apply the observational bound to the absolute value of our result:\footnote{We argue that the $95\%$ C.L. can be applied to the CnC contribution to $h_{c,\rm eq}$ because if this contribution was larger in magnitude than the sum total, positive contributions to $h_{c,\rm eq}$ would have to dominate at higher and lower frequencies to prevent the total from becoming negative at large or small $\omega$. But this requires a strong fine-tuning so that these contributions are minimal exactly at the observed frequency. We assume the probability of such a coincidence to be negligible.}
\be
\left|(h_{c,{\rm eq}})^{(X)}(2\pi\times 0.195 ~{\rm yr}^{-1})\right| < 3.20\times 10^{-15} ~~(95\%~{\rm C.L.}).
\ee
This leads to constraints on the masses of particles of different species $X=\phi,\psi$, etc., listed in Table \ref{table}.

\begin{table}[ht!]
\begin{center}
\begin{tabular}{|c|p{3.cm}|}
\multicolumn{2}{c} {\bf Beyond-SM Mass Bounds } \\
\hline
\textit{Particle Species} & \textit{$2\sigma$ upper bound}  \\
\hline 
Real Scalar \ \ \ & $m<600$ GeV \\
\hline 
Complex Scalar \ \ \ & $m<525$ GeV  \\
\hline 
Majorana \ \ \ & $m<525$ GeV  \\
\hline 
Dirac \ \ \ & $m<450$ GeV  \\
\hline 
Real Vector \ \ \ & $m<480$ GeV  \\
\hline 
Complex Vector \ \ \ & $m<420$ GeV \\
\hline
\end{tabular}
\end{center}
\caption{Mass bounds on Beyond Standard Model particles of different species, from PSR J1909-3744.}
\label{table}
\end{table}

\begin{figure}
\begin{center}
\hspace{-0.3cm}
\includegraphics[scale=0.43]{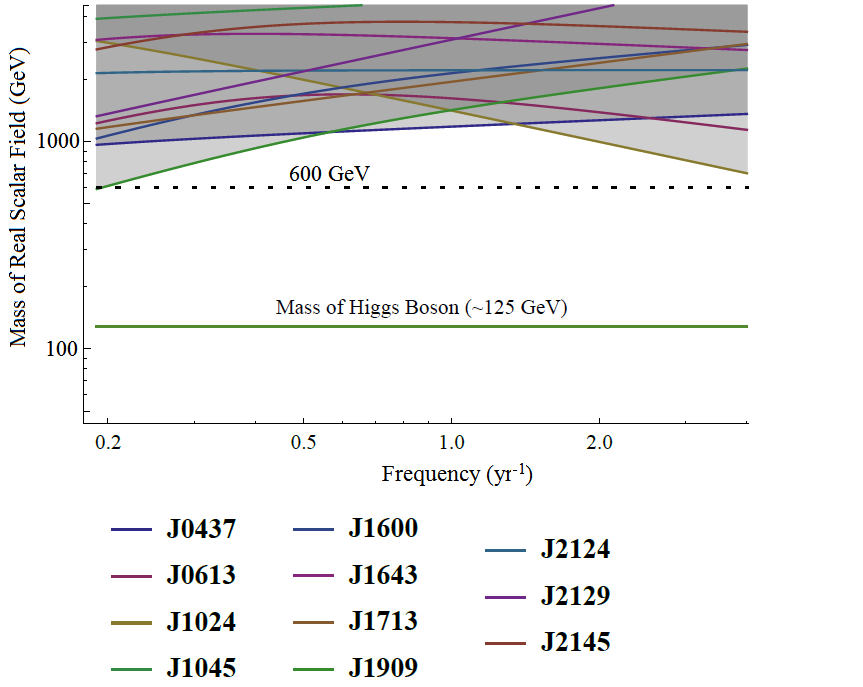}
\end{center}
\caption{Mass bounds on Beyond Standard Model (BSM) real scalar fields, based on pulsar timing noise models of 11 millisecond pulsars \cite{Reardon:2015kba}. The colored lines show constraints on the amplitude of fluctuations in the period of pulses emitted by various pulsars, converted to scalar field mass using Eq. \eqref{pulsar_CnC}. The strongest constraint comes from PSR J1909-3744 at $f\approx0.2 \ \text{yr}^{-1}$.  Mass bounds on higher-spin particles are slightly stronger (see Table \ref{table}).}
\label{fig:mass_bounds}
\end{figure}

Figure \ref{fig:mass_bounds} compares the resulting mass bounds inferred from the timing noise models of 11 millisecond pulsars \cite{Reardon:2015kba}, showing that the strongest bound indeed comes from PSR J1909-3744.

\textit{Discussion.} We should note that the negative sign of the power spectrum, as found here, often happens in quantum field theories due to renormalization of UV divergences \cite{Hsiang:2010uv}. This does force us to conjecture an unknown \textit{positive} contribution from UV and/or pulsar physics, which would guarantee the positivity of any observable quantity. Admittedly, this is a less than satisfactory aspect of our proposal, but we are forced to it by interpreting the low-frequency part of stress tensor fluctuations (which is the only UV-convergent part in real space) as a genuine contribution to gravitating stress-energy. Whether there exists a compelling argument for such a conjectured consistency condition remains to be seen, and we leave it for future work.

We could also imagine a fine-tuning of the UV contour integrals $\C_\infty^\pm$ to cancel the IR observables. For example, such a cancellation could have been enforced by the Poincare symmetry of Minkowski space. However, given that we do not live in Minkowski space, we do not see any technically natural mechanism to impose this cancellation.  

Lastly, one might wonder about the role of matter interactions, which could lead to, e.g., instability of massive particles. While we have only considered free fields, as long as matter interactions are weak (with small loop corrections), we do not expect them to significantly alter our result.

\textit{Summary.} We have computed the contribution to pulsar timing noise due to pulse propagation through metric fluctuations sourced by the vacuum stress-energy of quantum fields. We found that for a field of mass $m$, there is a finite low-frequency component (observable at $\omega\lesssim{\rm yr}^{-1}$) along with UV-divergent high-frequency terms ($\omega \gtrsim m$). Current constraints on pulsar timing noise requires that Beyond Standard Model fields have masses less than 600 GeV $\simeq 4 \times$ top quark mass. The top quark (as the heaviest SM particle) will source the leading SM contributions to low-frequency vacuum stress-energy and hence to pulsar timing noise, so we predict a contribution to pulsar timing noise given by Eq. \eqref{pulsar_CnC} with $c_X=c_{\psi,\rm Dirac}$ and $m\approx 170$ GeV. Therefore, a factor of $\sim$ 10 improvement on $h_{c,{\rm eff}}$ measurement, expected over the next 10-15 years,  will bring the pulsar timing noise sensitivity close to ruling out (or confirming) our prediction. Competitive bounds may also come from LIGO or LISA gravitational wave observatories, as well as the angular structure of the pulsar timing noise \cite{Hellings:1983fr}, but we shall defer a detailed analysis to future work. 

To conclude, it is hard to overstate the significance of our finding for current and upcoming searches for BSM physics, specifically  in particle colliders. While one can dismiss our bounds as a possible artifact of a UV regularization scheme, given the enormous stakes in play (e.g., for planning of a 100-TeV collider \cite{Arkani-Hamed:2015vfh}), and the fleeting habits of BSM signals (e.g., the 750 GeV diphoton excess \cite{Aaboud:2016tru,Khachatryan:2016hje}), dismissal may not be the wisest course of action!

{\it Acknowledgement:} We would like to thank Cliff Burgess, William Donnelly, and Richard Woodard for useful discussions. This work has been partially supported by Natural Sciences and Engineering Research Council of Canada (NSERC), University of Waterloo, and Perimeter Institute for Theoretical Physics (PI). Research at PI is supported by the 
Government of Canada through the Department of Innovation, Science and Economic Development Canada and by the Province of Ontario through 
the Ministry of Research, Innovation and Science.

\bibliographystyle{apsrev4-1}
\bibliography{CnC_pulsar}

\end{document}